\documentclass{stacs_proc}

\usepackage{xspace,amstext,amssymb,url}
\usepackage{amsmath}
\usepackage{graphicx,color,epsfig}
\usepackage{subfigure}
\usepackage{enumerate}

\newcommand{\nzc}{\newcommand}
\nzc{\shuf}{\&}
\nzc{\shb}{\mathbin\&}
\nzc{\shuffle}{\shuf}
\nzc{\mycount}{\#}
\nzc{\re}{\text{RE}\xspace}
\nzc{\rein}{\text{RE($\cap$)}\xspace}
\nzc{\recom}{\text{RE($\neg$)}\xspace}
\nzc{\reincom}{\text{RE($\cap$,$\neg$)}\xspace}
\nzc{\res}{\text{REs}\xspace}
\nzc{\rec}{\text{RE($\mycount$)}\xspace}
\nzc{\recs}{\text{RE($\mycount$)}s\xspace}
\nzc{\rei}{\text{RE($\shuffle$)}\xspace}
\nzc{\reis}{\text{RE($\shuffle$)}s\xspace}
\nzc{\reci}{\text{RE($\mycount,\shuf$)}\xspace}
\nzc{\recis}{\text{RE($\mycount,\shuf$)s}\xspace}
\nzc{\chare}{\text{CHARE}\xspace}
\nzc{\charec}{\text{CHARE($\mycount$)}\xspace}
\nzc{\lin}{\text{RE(linear)}\xspace}
\nzc{\strongLin}{\text{RE(strong-linear)}\xspace}
\nzc{\disjoint}{\ensuremath \text{$^{\text{disj}}$}\xspace}
\nzc{\strongLinDis}{\ensuremath \text{\strongLin$^{\text{disj}}$} \xspace}

\nzc{\sym}{\text{Sym}\xspace}

\nzc{\configuration}{\text{configuration}\xspace}
\nzc{\configurations}{\text{configurations}\xspace}

\nzc{\nta}{\text{NTA}\xspace}
\nzc{\nfa}{\text{NFA}\xspace}
\nzc{\nfas}{\text{NFAs}\xspace}
\nzc{\nfaci}{\text{NFA$(\mycount,\shuffle)$}\xspace}
\nzc{\nfacis}{\text{NFA$(\mycount,\shuffle)$s}\xspace}
\nzc{\nfac}{\text{NFA$(\mycount)$}\xspace}
\nzc{\nfacs}{\text{NFA$(\mycount)$s}\xspace}
\nzc{\nfai}{\text{NFA$(\shuffle)$}\xspace}
\nzc{\nfais}{\text{NFA$(\shuffle)$s}\xspace}

\nzc{\patree}{\ensuremath \text{$(\re,R)_{\exists}$}\xspace}
\nzc{\patreu}{\ensuremath \text{$(\re,R)_{\forall}$}\xspace}
\nzc{\patlineaire}{\ensuremath \text{$(lineair,R)_{\exists}$}\xspace}
\nzc{\patlineairu}{\ensuremath \text{$(lineair,R)_{\forall}$}\xspace}
\nzc{\patstronge}{\ensuremath \text{$(stronglineair(\cap = \emptyset),R)_{\exists}$}\xspace}
\nzc{\patstrongu}{\ensuremath \text{$(stronglineair(\cap = \emptyset),R)_{\forall}$}\xspace}

\nzc{\nat}{\Bbb{N}}
\nzc{\Dom}{\text{Dom}}
\nzc{\lab}{\text{lab}}
\nzc{\depth}{\text{depth}}

\nzc{\logspace}{{\sc logspace}\xspace} 
\nzc{\nlogspace}{{\sc nlogspace}\xspace} 
\nzc{\ptime}{{\sc ptime}\xspace} 
\nzc{\np}{{\sc np}\xspace} 
\nzc{\conp} {\textrm{co}\textsc{np}\xspace}
\nzc{\pspace}{{\sc pspace}\xspace} 
\nzc{\exptime}{{\sc exptime}\xspace}
\nzc{\expspace}{{\sc expspace}\xspace}
\nzc{\nexpspace}{{\sc nexpspace}\xspace}
\nzc{\nexptime}{{\sc nexptime}\xspace}
\nzc{\blogspace}{{\scriptsize {\bf LOGSPACE}}}
\nzc{\bnlogspace}{{\scriptsize {\bf  NLOGSPACE}}}
\nzc{\bptime}{{\scriptsize {\bf PTIME}}}
\nzc{\bnp}{{\scriptsize {\bf NP}}}
\nzc{\bconp}{{\scriptsize {\bf coNP}}}
\nzc{\bpspace}{{\scriptsize {\bf PSPACE}}}
\nzc{\bexptime}{{\scriptsize {\bf EXPTIME}}}
\nzc{\bnexptime}{{\scriptsize {\bf NEXPTIME}}}
\nzc{\bexpspace}{{\scriptsize {\bf EXPSPACE}}}

\nzc{\expo}{\text{exp}\xspace}

\nzc{\cO}{{\mathcal O}}
\nzc{\cR}{{\cal R}\xspace}
\nzc{\cS}{{\mathcal S}}
\nzc{\cT}{{\mathcal T}}
\nzc{\cM}{{\mathcal M}}
\nzc{\subs}{\subseteq}
\renewcommand{\epsilon}{\varepsilon}

\nzc{\first}{\text{First}\xspace}
\nzc{\last}{\text{Last}\xspace}
\nzc{\follow}{\text{Follow}\xspace}
\nzc{\nfirst}{\text{Not-First}\xspace}
\nzc{\nfollow}{\text{Not-Follow}\xspace}

\nzc{\uab}{\text{RE(1UA)}}
\nzc{\uabc}{\text{RE($\mycount$)-UAB}}
\nzc{\uabw}{\ensuremath \text{UAB}_W}
\nzc{\uabwc}{\ensuremath \text{UAB}^{\#}_W\xspace}
\nzc{\uabs}{\ensuremath \text{UAB}_S}
\nzc{\uabsc}{\ensuremath \text{UAB}^{\#}_S\xspace}

\nzc{\satisfiability}{{\sc satisfiability}\xspace}
\nzc{\inclusion}{{\sc inclusion}\xspace}
\nzc{\intersection}{{\sc intersection}\xspace}
\nzc{\equivalence}{{\sc equivalence}\xspace}
\nzc{\emptiness}{{\sc emptiness}\xspace}
\nzc{\membership}{{\sc membership}\xspace}
\nzc{\simplification}{{\sc simplification}\xspace}
\nzc{\reachability}{{\sc reachability}\xspace}
\nzc{\onerun}{{\sc one run}\xspace}
\nzc{\partition}{{\sc partition}\xspace}
\nzc{\universality}{{\sc universality}\xspace}

\nzc{\edtd}{\text{EDTD}\xspace}
\nzc{\edtds}{\text{EDTDs}\xspace}
\nzc{\edtdre}{\text{\edtd}\xspace}
\nzc{\edtdc}{\text{\edtd}($\mycount$)\xspace}
\nzc{\edtdi}{\text{\edtd}($\shuffle$)\xspace}
\nzc{\edtdci}{\text{\edtd}($\mycount$,$\shuffle$)\xspace}
\nzc{\dtd}{\text{DTD}\xspace}
\nzc{\dtdc}{\text{\dtd}($\mycount$)\xspace}
\nzc{\dtdi}{\text{\dtd}($\shuffle$)\xspace}
\nzc{\dtdci}{\text{\dtd}($\mycount$,$\shuffle$)\xspace}

\nzc{\suf}{\text{Suffix}\xspace}
\nzc{\prefix}{\text{Prefix}\xspace}
\nzc{\enc}{\text{enc}\xspace}
\nzc{\construct}{\text{construct}\xspace}
\nzc{\init}{\text{init}\xspace}

\renewcommand{\clubsuit}{\triangle}
\renewcommand{\spadesuit}{\triangle}

\newcommand{\triangleOne}{\rhd}
\newcommand{\triangleTwo}{\triangle}
\newcommand{\In}{\text{In}}
\newcommand{\Out}{\text{Out}}

\theoremstyle{plain}
\newtheorem{mytheorem}[thm]{Theorem}

\newtheorem{mylemma}[thm]{Lemma}
\theoremstyle{definition}
\newtheorem{mydefinition}[thm]{Definition}{\bfseries}{\rm}
{\itshape}{\itshape}
{\itshape}{\itshape}

\newenvironment{myproof}{\begin{proof}}{\end{proof}}

\begin{document}

\title[Succinctness of regular expressions]{Succinctness of the
  Complement and Intersection of Regular Expressions}

\author{W. Gelade}{Wouter Gelade}
\address{Hasselt University and Transnational University of Limburg, School for Information Technology}  
\email{firstname.lastname@uhasselt.be}  

\author{F. Neven}{Frank Neven}

\thanks{Wouter Gelade is a Research Assistant of the Fund for
  Scientific Research - Flanders (Belgium)}

  \begin{abstract}
    We study the succinctness of the complement and intersection
    of 
    regular expressions. In particular, we show that when constructing a
    regular expression defining the complement of a given regular
    expression, a double exponential size increase cannot be avoided.
    Similarly, when constructing a regular expression defining the
    intersection of a fixed and an arbitrary number of regular
    expressions, an exponential and double exponential size increase,
    respectively, can in worst-case not be avoided. All mentioned lower
    bounds improve the existing ones by one exponential and are tight in
    the sense that the target expression can be constructed in the
    corresponding time class, i.e., exponential or double exponential
    time. 
    As a by-product, we generalize a theorem by Ehrenfeucht and Zeiger
    stating that there is a class of DFAs which are exponentially more
    succinct than regular expressions, to a fixed four-letter alphabet.
    When the given regular expressions are one-unambiguous, as for
    instance required by the XML Schema specification, the complement
    can be computed in polynomial time whereas the bounds concerning
    intersection continue to hold.  For the subclass of
    single-occurrence regular expressions, we prove a tight exponential
    lower bound for intersection.
  \end{abstract}

\maketitle

\stacsheading{2008}{325-336}{Bordeaux}
\firstpageno{325}

\section{Introduction}

The two central questions addressed in this paper are the following.
Given regular expressions $r,r_1,\ldots,r_k$ over an alphabet $\Sigma$,
\begin{enumerate}
\item what is the complexity of constructing a regular expression $r_\neg$
  defining $\Sigma^*\setminus L(r)$, that is, the complement of $r$?

\item what is the complexity of constructing a regular
expression $r_\cap$  defining $L(r_1) \cap \cdots \cap L(r_k)$?
\end{enumerate}
In both cases, the naive algorithm takes time double exponential in
the size of the input. Indeed, for the complement, transform $r$ to an
NFA and determinize it (first exponential step), complement it and
translate back to a regular expression (second exponential step). For
the intersection there is a similar algorithm through a translation to
NFAs, taking the crossproduct and a retranslation to a regular
expression. Note that both algorithms do not only take double
exponential time but also result in a regular expression of double
exponential size.  In this paper, we exhibit classes of regular
expressions for which this double exponential size increase cannot be
avoided. Furthermore, when the number $k$ of regular expressions is
fixed, $r_\cap$ can be constructed in exponential time and we prove a
matching lower bound for the size increase. In addition, we consider
the fragments of one-unambiguous and single-occurrence regular
expressions relevant to XML schema
languages~\cite{myvldb,vldb2007,ghelli,tods2006}. Our main results are
summarized in Table~\ref{tab:overview}.

The main technical part of the paper is centered around the
generalization of a result by Ehrenfeucht and
Zeiger~\cite{DBLP:journals/jcss/EhrenfeuchtZ76}. They exhibit a class
of languages $(Z_n)_{n\in\nat}$ each of which can be accepted by a DFA
of size $\cO(n^2)$ but cannot be defined by a regular expression of
size smaller than $2^{n-1}$. The most direct way to define $Z_n$ is by
the DFA that accepts it: the DFA is a graph consisting of $n$ states,
labeled 0 to $n - 1$, which are fully connected and the edge between
state $i$ and $j$ carries the label $a_{i,j}$. It now accepts all
paths in the graph, that is, all strings of the form
$a_{i_0,i_1}a_{i_1,i_2} \cdots a_{i_k,i_{k+1}}$. Note that the
alphabet over which $Z_n$ is defined grows quadratically with $n$.  We
generalize their result to a four-letter alphabet. In particular, we
define $K_n$ as the binary encoding of $Z_n$ using a suitable encoding
for $a_{i,j}$ and prove that every regular expression defining $K_n$
should be at least of size $2^n$.  As integers are encoded in binary
the complement and intersection of regular expressions can now be used
to separately encode $K_{2^n}$ (and slight variations thereof) leading
to the desired results. In \cite{ShallitRegEx} the same generalization
as obtained here is attributed to Waizenegger~\cite{waizenegger}.
Unfortunately, we believe that proof to be incorrect as we discuss in
the full version of this paper. 

Although the succinctness of various automata
models have been investigated in depth~\cite{globerman_harel} and more
recently those of logics over (unary alphabet)
strings~\cite{DBLP:journals/lmcs/GroheS05}, the succinctness of
regular expressions has hardly been addressed. For the complement of a
regular expression an exponential lower bound is given by Ellul et
al~\cite{ShallitRegEx}. For the intersection of an arbitrary number of
regular expressions Petersen gave an exponential lower
bound~\cite{DBLP:conf/stacs/Petersen02}, while Ellul et
al~\cite{ShallitRegEx} mention a quadratic lower bound for the
intersection of two regular expressions.  In fact, in
\cite{ShallitRegEx}, it is explicitly asked what the maximum
achievable blow-up is for the complement of one and the intersection
of two regular expressions (Open Problems 4 and 5).  Although we do not
answer these questions in the most precise way, our lower bounds
improve the existing ones by one exponential and are tight in the
sense that the target expression can be constructed in the time class
matching the space complexity of the lower bounds.

Succinctness of complement and intersection relate to the succinctness
of semi-extended (\rein) and extended regular expressions
(\reincom).  These are regular expressions augmented with
intersection and both complement and intersection operators,
respectively. Their membership problem has been extensively studied
\cite{DBLP:journals/ipl/JiangR91,DBLP:conf/mfcs/KupfermanZ02,Myers92,DBLP:conf/stacs/Petersen02,DBLP:conf/rta/RouV03}. 
Furthermore, non-emptiness and equivalence of \reincom 
is non-elementary \cite{DBLP:conf/stoc/StockmeyerM73}. For
\rein, 
 inequivalence is \expspace-complete
\cite{DBLP:conf/icalp/Furer80,Hunt,DBLP:journals/ipl/Robson79}, and
non-emptiness is \pspace-complete \cite{DBLP:conf/icalp/Furer80,Hunt}
even when restricted to the intersection of a (non-constant) number of
regular expressions \cite{DBLP:conf/focs/Kozen77}. Several of these
papers hint upon the succinctness of the intersection operator and
provide dedicated techniques in dealing with the new operator directly
rather than through a translation to ordinary regular
expressions~\cite{DBLP:conf/mfcs/KupfermanZ02,DBLP:conf/stacs/Petersen02}.
Our results present a double exponential lower bound in translating
\rein to \re and therefore justify even more the development for
specialized techniques.

A final motivation for this research stems from its application in the
emerging area of
XML-theory~\cite{libkinICALP2005,nevenCSL2002,SchwentickJCSS,vianuSTACS2003}.  
From a formal language viewpoint, XML documents can be seen as labeled
unranked trees and collections of these documents are defined by
schemas. A schema can take various forms, but the most common ones are
Document Type Definitions (DTDs)~\cite{w3} and XML Schema Definitions
(XSDs)~\cite{w3schema} which are grammar based formalisms with regular
expressions at right-hand sides of
rules~\cite{tods2006,DBLP:journals/toit/MurataLMK05}.  Many questions
concerning schemas reduce to corresponding questions on the classes of
regular expressions used as right-hand sides of rules as is
exemplified for the basic decision problems studied in
\cite{DBLP:conf/icdt/GeladeMN07} and
\cite{DBLP:conf/mfcs/MartensNS04}. Furthermore, the lower bounds
presented here are utilized in~\cite{geladenevendbpl} to prove, among
other things, lower bounds on the succinctness of existential and
universal pattern-based schemas on the one hand, and single-type EDTDs
(a formalization of XSDs) and DTDs, on the other hand.
As the DTD and XML Schema specification require regular expressions
occurring in rules to be \emph{deterministic}, formalized by
Br\"uggemann-Klein and Wood in terms of one-unambiguous regular
expressions~\cite{Bruggemann-KleinW1998-dup}, we also investigate the
complement and intersection of those. In particular, we show that a
one-unambiguous regular expressions can be complemented in polynomial
time, whereas the lower bounds concerning intersection carry over from
unrestricted regular expressions. A study in \cite{myvldb} reveals
that most of the one-unambiguous regular expression used in practice
take a very simple form: every alphabet symbol occurs at most once. We
refer to those as single-occurrence regular expressions (SOREs) and
show a tight exponential lower bound for intersection.


\begin{table}
  \centering
  \begin{tabular}{|l|c|c|c|}
    \hline
     & complement & intersection (fixed) & intersection
     (arbitrary)\\\hline \hline
   regular expression & 2-exp & exp & 2-exp\\\hline
   one-unambiguous & poly & exp & 2-exp\\\hline
   single-occurrence & poly & exp & exp\\\hline
  \end{tabular}
  \caption{Overview of the size increase for the various operators and
subclasses. All non-polynomial complexities are tight.}
  \label{tab:overview}
\end{table}



\smallskip
\noindent
{\bf Outline.} 
In Section~2, we introduce the necessary notions concerning
(one-unambiguous) regular expressions and automata.  In Section~3, we
extend the result by Ehrenfeucht and Zeiger to a fixed alphabet using
the family of languages $(K_n)_{n\in\nat}$.  In Section~4, we consider
the succinctness of complement.  In Section~5, we consider the
succinctness of intersection of several classes of regular
expressions.  We conclude in Section~6. A version of this paper
containing all proofs is available from the authors' webpages.

\section{Preliminaries}


\subsection{Regular expressions}

By $\nat$ we denote the natural numbers without zero.
For the rest of the paper, $\Sigma$ always denotes a finite alphabet.
A \emph{$\Sigma$-string} (or simply string) is a finite sequence $w =
a_1\cdots a_n$ of $\Sigma$-symbols. We define the length of $w$,
denoted by $|w|$, to be $n$.  We denote the empty string by
$\varepsilon$.  The set of \emph{positions of $w$} is $\{1,\ldots,n\}$
and the \emph{symbol of $w$ at position $i$} is $a_i$.  By $w_1 \cdot
w_2$ we denote the \emph{concatenation} of two strings $w_1$ and
$w_2$.  As usual, for readability, we denote the concatenation of
$w_1$ and $w_2$ by $w_1w_2$. The set of all strings is denoted by
$\Sigma^*$ and the set of all non-empty strings by $\Sigma^+$. A
\emph{string language} is a subset of $\Sigma^*$. For two string
languages $L,L'\subseteq \Sigma^*$, we define their concatenation $L
\cdot L'$ to be the set $\{w \cdot w'\mid w\in L, w'\in L'\}$. We
abbreviate $L \cdot L \cdots L$ ($i$ times) by $L^i$.

The set of \emph{regular expressions} over $\Sigma$, denoted by \re,
is defined in the usual way: $\emptyset$, $\varepsilon$, and every
$\Sigma$-symbol is a regular expression; and when $r_1$ and $r_2$ are
regular expressions, then $r_1\cdot r_2$, $r_1 + r_2$, and $r_1^*$ are
also regular expressions. 

By $\reincom$ we denote the class of \emph{extended regular expressions},
that is, \re extended with intersection and complementation
operators. So, when $r_1$ and $r_2$ are \reincom-expressions then so
are $r_1 \cap r_2$ and $\neg r_1$. By \rein and \recom we denote RE
extended solely with the intersection and complement operator,
respectively.

The language defined by an extended regular expression $r$, denoted by $L(r)$,
is inductively defined as follows: $L(\emptyset) = \emptyset$;
$L(\varepsilon)=\{\varepsilon\}$; $L(a)=\{a\}$; $L(r_1r_2)=L(r_1)\cdot
L(r_2)$; $L(r_1 + r_2)=L(r_1) \cup L(r_2)$;
$L(r^*)=\{\varepsilon\}\cup \bigcup_{i=1}^\infty L(r)^i$; $L(r_1 \cap
r_2) = L(r_1) \cap L(r_2)$; and $L(\neg r_1) = \Sigma^* \setminus
L(r_1)$. 

By $\bigcup_{i=1}^k r_i$, and $r^k$, with $k \in \nat$,
we abbreviate the expression $r_1+\cdots + r_k$,
and $rr\cdots r$ ($k$-times), respectively. For a set $S = \{a_1,
\ldots, a_n\} \subseteq \Sigma$, we abbreviate by $S$ the regular
expression $a_1 + \cdots + a_n$.

We define the \emph{size} of an extended regular expression $r$ over $\Sigma$,
denoted by $|r|$, as the number of $\Sigma$-symbols and operators
occurring in $r$ disregarding parentheses. This is equivalent to
the length of its (parenthesis-free) reverse 
Polish form \cite{DBLP:journals/tcs/Ziadi96}.
Formally, $|\emptyset|=|\varepsilon|=|a|=1$, for
$a\in\Sigma$, $|r_1r_2|=|r_1\cap r_2|=|r_1+ r_2|=|r_1|+|r_2|+1$, and
$|\neg r|=|r^*|=|r|+1$.

Other possibilities considered in the literature for defining the size
of a regular expression are: (1) counting all symbols, operators, and
parentheses \cite{AhoHopcroftUllman,DBLP:conf/mfcs/IlieY02}; or, (2)
counting only the $\Sigma$-symbols. However, Ellul et
al.~\cite{ShallitRegEx} have shown that for regular expressions (so,
without $\neg$ and $\cap$), provided they are preprocessed by
syntactically eliminating superfluous $\emptyset$- and
$\varepsilon$-symbols, and nested stars, the three length measures are
identical up to a constant multiplicative factor. For extended regular
expressions, counting only the $\Sigma$-symbols is not sufficient,
since for instance the expression $(\neg \varepsilon)(\neg
\varepsilon)(\neg \varepsilon)$ does not contain any
$\Sigma$-symbols. Therefore, we define the size of an expression as
the length of its reverse Polish form.

\subsection{One-unambiguous regular expressions and SOREs}

As mentioned in the introduction, several XML schema languages
restrict regular expressions occurring in rules to be
\emph{deterministic}, formalized by Br\"uggemann-Klein and
Wood~\cite{Bruggemann-KleinW1998-dup} in terms of one-unambiguity.
We introduce this notion next.

To indicate different occurrences of the same symbol in a regular expression, we
mark symbols with subscripts. For instance, the \emph{marking} of $(a +
b)^*a + bc$ is $(a_1 + b_2)^*a_3 + b_4c_5$. We denote 
 by $r^\flat$ the marking of $r$ and by $\sym(r^\flat)$ the subscripted symbols
 occurring in $r^\flat$. When $r$ is a marked expression, then
$r^\natural$ over $\Sigma$ is obtained from $r$ by dropping all
subscripts. This notion is extended to words and
languages in the usual way.

\begin{mydefinition}
  \rm
A regular expression $r$ is \emph{one-unambiguous} iff for
all strings $w,u,v \in \sym(r^{\flat})^*$, and all symbols $x,y \in
\sym(r^\flat)$, the conditions $uxv, uyw \in L(r^\flat)$ and $x \neq
y$ imply $x^\natural \neq y^\natural$. 
\end{mydefinition}

For instance, the regular expression $r = a^*a$, with marking $r^\flat
= a_1^*a_2$, is not one-unambiguous. Indeed, the marked strings
$a_1a_2$ and $a_1a_1a_2$ both in $L(r^\flat)$ do not satisfy the
conditions in the previous definition. The equivalent expression
$aa^*$, however, is one-unambiguous.  The intuition behind the
definition is that positions in the input string can be matched in a
deterministic way against a one-unambiguous regular expression without
looking ahead. 
For instance, for the expression $aa^*$, the first $a$ of
an input string is always matched against the leading $a$ in the
expression, while every subsequent $a$ is matched against the last
$a$. Unfortunately, one-unambiguous regular languages do not form a
very robust class as they are not even closed under the Boolean
operations~\cite{Bruggemann-KleinW1998-dup}.

The following subclass captures the class of regular expressions
occurring in XML schemas on the Web~\cite{myvldb}:
\begin{mydefinition}
  \rm A \emph{single-occurrence regular expression (SORE)} is a
  regular expression where every alphabet symbol occurs at most
  once. In addition, we allow the operator $r^+$ which defines $rr^*$.
\end{mydefinition}
For instance, $(a+b)^+c$ is a SORE while $a^*(a+b)^+$ is not.
Clearly, every SORE is one-unambiguous. Note that SOREs define local
languages and that over a fixed alphabet there are only finitely many
of them.


  

\subsection{Finite automata}

A non-deterministic finite automaton (NFA) $A$ is a 4-tuple
$(Q,q_0,\delta,F)$ where $Q$ is the set of states, $q_0$ is the
initial state, $F$ is the set of final states and $\delta \subseteq Q
\times \Sigma \times Q$ is the transition relation. We write $q
\Rightarrow_{A,w} q'$ when $w$ takes $A$ from state $q$ to $q'$. So,
$w$ is accepted by $A$ if $q_0 \Rightarrow_{A,w} q'$ for some $q' \in
F$. The set of strings accepted by $A$ is denoted by $L(A)$. The size
of an NFA is $|Q| + |\delta|$. An NFA is \emph{deterministic} (or a
DFA) if for all $a \in \Sigma, q \in Q$, $|\{(q,a,q') \in \delta \mid
q' \in Q\}| \leq 1$.

We make use of the following known results.
\begin{mytheorem} \label{theo:regular-operations}
  Let $A_1,\ldots,A_m$ be NFAs over $\Sigma$ with $|A_i| =
  n_i$
for $i\leq m$, and $|\Sigma| = k$.
  \begin{enumerate}
  \item \label{theo:regular-operations-1} A regular expression $r$,
    with $L(r) = L(A_1)$, can be constructed in time
    $\cO(m_1k4^{m_1})$, where $m_1$ is the number of states of $A_1$~\cite{McNaughton60,ShallitRegEx}.
  \item \label{theo:regular-operations-4} A DFA $B$ with $2^{n_1}$
    states, such that $L(B) =
    L(A_1)$, can be constructed in time $\cO(2^{n_1})$ \cite{Yu}.
  \item \label{theo:regular-operations-5} A DFA $B$ with $2^{n_1}$
    states, such that $L(B) =
    \Sigma^* \setminus L(A_1)$, can be constructed in time
    $\cO(2^{n_1})$ \cite{Yu}.
  \item \label{theo:regular-operations-7} Let $r \in \re$. An NFA $B$
    with $|r|+1$ states,
    such that $L(B) = L(r)$, can be constructed in time
    ${\cO}(|r|\cdot|\Sigma|)$ \cite{DBLP:journals/tcs/Bruggemann-Klein93}.
  \item \label{theo:regular-operations-6} Let $r \in \rein$. An NFA
    $B$ with $2^{|r|}$ states, such that $L(B) = L(r)$, can be constructed
    in time exponential in the size of $r$
    \cite{DBLP:conf/icalp/Furer80}.
  \end{enumerate}
\end{mytheorem}

\section{A generalization of a Theorem by Ehrenfeucht and Zeiger to a
  fixed alphabet}
\label{sec:language}

We first introduce the family $(Z_n)_{n\in\nat}$ defined by
Ehrenfeucht and Zeiger over an alphabet whose size grows
quadratically with the parameter
$n$~\cite{DBLP:journals/jcss/EhrenfeuchtZ76}:
\begin{mydefinition}\label{def:ehr}
  Let $n \in \nat$ and $\Sigma_n = \{a_{i,j}\mid 0 \leq i,j \leq
  n-1\}$. Then, $Z_n$ contains exactly all strings of the form
  $a_{i_0,i_1}a_{i_1,i_2}\cdots a_{i_{k-1},i_k}$ where $k \in \nat$.
\end{mydefinition}
A way to interpret $Z_n$ is to consider the DFA with states
$\{0,\ldots,n-1\}$ which is fully connected and where the edge between
state $i$ and $j$ is labeled with $a_{i,j}$.  The language $Z_n$ then
consists of all paths in the DFA.~\footnote{Actually, in
\cite{DBLP:journals/jcss/EhrenfeuchtZ76}, only paths from state 0 to
state $n-1$ are considered. We use our slightly modified definition as
it will be easier to generalize to a fixed arity alphabet suited for
our purpose in the sequel.}

Ehrenfeucht and Zeiger obtained the succinctness of DFAs
with respect to regular expressions through the following theorem:
\begin{mytheorem}[\cite{DBLP:journals/jcss/EhrenfeuchtZ76}] \label{theo:ehrenfeucht}
For $n\in \nat$, any regular expression defining $Z_n$ must
be of size at least $2^{n-1}$. Furthermore, there is a DFA
 of size ${\cO}(n^2)$ accepting $Z_n$.
\end{mytheorem}

Our language $K_n$ is then the straightforward binary encoding of
$Z_n$ that additionally swaps the pair of indices in every symbol
$a_{i,j}$.  Thereto, for $a_{i,j} \in \Sigma_n$, define the function
$\rho_n$ as $$\rho_n(a_{i,j}) = \enc(j)\$\enc(i)\#,$$ where $\enc(i)$
and $\enc(j)$ denote the $\lceil \log(n) \rceil$-bit binary encodings
of $i$ and $j$, respectively. Note that since $i, j < n$, $i$ and $j$
can be encoded using only $\lceil \log(n) \rceil$-bits. We extend the
definition of $\rho_n$ to strings in the usual way: $\rho_n(
a_{i_0,i_1}\cdots a_{i_{k-1},i_k})=
\rho_n(a_{i_0,i_1})\cdots \rho_n(a_{i_{k-1},i_k})$.

We are now ready to define $K_n$.
\begin{mydefinition} \label{def:K} Let $\Sigma_K =
  \{0,1,\$,\#\}$. For $n\in\nat$, let $K_n = \{\rho_n(w) \mid w \in
  Z_n\}$.
\end{mydefinition}
For instance, for $n = 5$, $w =
a_{3,2}a_{2,1}a_{1,4}a_{4,2} \in Z_5$ and thus $$\rho_n(w) =
010\$011\#001\$010\#100\$001\#010\$100\# \in K_5.$$

We generalize
the previous theorem as follows:
\begin{mytheorem} \label{lem:size-reg-fixed} For any $n \in \nat$,
  with $n \geq 2$, 
  \begin{enumerate}
  \item any regular expression defining $K_n$ is of size at least
    $2^{n}$; and,
  \item there is a DFA $A_n$ of size ${\cO}(n^2\log{n})$ defining
    $K_n$.
\end{enumerate}
\end{mytheorem}
The construction of $A_n$ is omitted. The rest of this
section is devoted to the proof of
Theorem~\ref{lem:size-reg-fixed}(1). It follows the structure of the
proof of Ehrenfeucht and Zeiger but is technically more involved as
it deals with binary encodings of integers.

We start by introducing some terminology. Let $w =
a_{i_0,i_1}a_{i_1,i_2}\cdots a_{i_{k-1},i_k}\in Z_n$. We say that
$i_0$ is the \emph{start-point} of $w$ and $i_k$ is its
\emph{end-point}.  Furthermore, we say that $w$ \emph{contains} $i$ or
$i$ \emph{occurs in} $w$ if $i$ occurs as an index of some symbol in
$w$. That is, $a_{i,j}$ or $a_{j,i}$ occurs in $w$ for some $j$.  For
instance, $a_{0,2}a_{2,2}a_{2,1} \in Z_5$, has start-point 0,
end-point 1, and contains 0, 1 and 2. 
The notions of contains, occurs, start- and end-point of
a string $w$ are also extended to $K_n$. So, the start and end-points
of $\rho_n(w)$ are the start and end-points of $w$, and $w$ contains
the same integers as $\rho_n(w)$.

  For a regular expression $r$, we say that $i$ is a \emph{sidekick}
  of $r$ when it occurs in every non-empty string defined by $r$.  A
  regular expression $s$ is a starred subexpression of a regular
  expression $r$ when $s$ is a subexpression of $r$ and is of the form
  $t^*$.




Now, the following lemma holds:
\begin{mylemma} \label{lem:always-contains} 
Any starred subexpression $s$ of a regular expression $r$ defining $K_n$
has a sidekick.
\end{mylemma}
We now say that a regular expression $r$ is \emph{normal} if every
starred subexpression of $r$ has a sidekick. In particular, any
expression defining $K_n$ is normal.  We say that a regular expression
$r$ \emph{covers} a string $w$ if there exist strings $u,u' \in
\Sigma^*$ such that $uwu' \in L(r)$. If there is a greatest integer
$m$ for which $r$ covers $w^m$, we call $m$ the \emph{index\/} of $w$ in $r$
and denote it by $I_w(r)$. In this case we say that $r$ is
\emph{$w$-finite}. Otherwise, we 
say that $r$ is \emph{$w$-infinite}. The index of a regular expression can be
used to give a lowerbound on its size according to the following
lemma.

  \begin{mylemma}[\cite{DBLP:journals/jcss/EhrenfeuchtZ76}] \label{lemma:index}
    For any regular expression $r$ and string $w$, if $r$ is
    $w$-finite, then $I_w(r) < 2|r|$.\footnote{In fact, in
      \cite{DBLP:journals/jcss/EhrenfeuchtZ76} the length of
      an expression is defined as the number of $\Sigma$-symbols
      occurring in it. However, since our length measure also contains
      these $\Sigma$-symbols, this lemma still holds in our setting.}
  \end{mylemma}

  Now, we can state the most important property of $K_n$.
  \begin{mylemma} \label{lem:size-induct} Let $n\geq 2$. For any
    $C\subseteq \{0, \ldots, n-1\}$ of cardinality $k$ and $i\in C$,
    there exists a string $w \in K_n$ with start- and end-point $i$
    only containing integers in $C$, such that any normal regular
    expression $r$ which covers $w$ is of size at least $2^k$.
  \end{mylemma}
  \begin{myproof} 
    The proof is by induction on the value of $k$. For $k = 1$, 
    $C = \{i\}$. Then, define $w =
    \enc(i)\$\enc(i)\#$, which satisfies all conditions and any
    expression covering $w$ must definitely have a size of at least 2.

    For the inductive step, let $C = \{j_1, \ldots, j_k\}$.  Define
    $C_\ell=C\setminus \{j_{(\ell \mod k)+1}\}$ and let $w_\ell$ be the
  string given by the induction hypothesis with respect to $C_\ell$ (of size
  $k-1$) and $j_\ell$. Note that $j_\ell\in C_\ell$.
 Further, define $m = 2^{k+1}$ and set 
  \begin{eqnarray*}
    \label{eq:1}
w =  \enc(j_1)\$\enc(i)\#w_1^m\enc(j_2)\$\enc(j_1)\#w_2^m
\enc(j_3)\$\enc(j_2)\#\cdots
  w_k^m\enc(i)\$\enc(j_k)\#.
\end{eqnarray*} 
Then, $w \in K_n$, has $i$ as start and end-point and only contains
integers in $C$. It only remains to show that any expression $r$ which
is normal and covers $w$ is of size at least $2^{k}$. 

Fix such a regular expression $r$.  If $r$ is $w_\ell$-finite for some
$\ell \leq k$. Then, $I_{w_\ell}(r_k) \geq m = 2^{k+1}$ by
construction of $w$. By Lemma~\ref{lemma:index}, $|r| \geq 2^k$ and we
are done.

Therefore, assume that $r$ is $w_\ell$-infinite for every $\ell \leq
k$. For every $\ell \leq k$, consider all subexpressions of $r$ which
are $w_\ell$-infinite.  It is easy to see that all minimal elements in
this set of subexpressions must be starred subexpressions.  Here and
in the following, we say that an expression is minimal with respect to a set
simply when no other expression in the set is a subexpression.
Indeed, a subexpression of the form $a$ or $\varepsilon$ can never be
$w_\ell$-infinite and a subexpression of the form $r_1r_2$ or $r_1 +
r_2$ can only be $w_\ell$-infinite if $r_1$ and/or $r_2$ are
$w_\ell$-infinite and is thus not minimal with respect to
$w_\ell$-infinity. Among these minimal starred subexpressions for
$w_\ell$, choose one and denote it by $s_\ell$. Let $E =
\{s_1,\ldots,s_k\}$. Note that since $r$ is normal, all its
subexpressions are also normal.  As in addition each $s_\ell$ covers
$w_\ell$, by the induction hypothesis the size of each $s_\ell$ is at
least $2^{k-1}$. Now, choose from $E$ some expression $s_\ell$ such
that $s_\ell$ is minimal with respect to the other elements in $E$.

As $r$ is normal and $s_\ell$ is a starred subexpression of $r$,
there is an integer $j$ such that every non-empty string in
$L(s_\ell)$ contains $j$. By definition of the strings $w_1, \ldots,
w_k$, there is some $w_p$, $p \leq k$, such that $w_p$ does not
contain $j$. Denote by $s_p$ the starred subexpression from $E$ which
is $w_p$-infinite. In particular, $s_\ell$ and $s_p$ cannot be the
same subexpression \mbox{of $r$}.

Now, there are three possibilities:
    \begin{itemize}
    \item $s_\ell$ and $s_p$ are completely disjoint subexpressions
      of $r$. That is, they are both not a subexpression of one
      another. By induction they must both be of size $2^{k - 1}$ and
      thus $|r| \geq 2^{k-1} + 2^{k-1} = 2^{k}$.

    \item $s_p$ is a strict subexpression of $s_\ell$. This is not possible
      since $s_\ell$ is chosen to be a minimum element from $E$.

      \item $s_\ell$ is a strict subexpression of $s_p$. We show that
	if we replace $s_\ell$ by
      $\varepsilon$ in $s_p$, then $s_p$ is still $w_p$-infinite. It
      then follows that $s_p$ still covers $w_p$, and thus $s_p$
      without $s_\ell$ is of size at least $2^{k-1}$. As $|s_\ell|
      \geq 2^{k-1}$ as well it follows that $|r| \geq 2^{k}$.

      To see that $s_p$ without $s_\ell$ is still $w_p$-infinite, recall
      that any non-empty string defined by $s_\ell$ contains $j$ and
      $j$ does not occur in $w_p$. Therefore, a full iteration of
      $s_\ell$ can never contribute to the matching of any number
      of repetitions of $w_p$. So, $s_p$ can only lose its
      $w_p$-infinity by this replacement if $s_\ell$ contains a
      subexpression which is itself $w_p$-infinite. However, this then
      also is a subexpression of $s_p$ and $s_p$ is chosen to be
      minimal with respect to $w_p$-infinity, a contradiction. We can
      only conclude that $s_p$ without $s_\ell$ is still
      $w_p$-infinite.


    \end{itemize}  \end{myproof}

  Since by Lemma~\ref{lem:always-contains} any expression defining
  $K_n$ is normal, Theorem~\ref{lem:size-reg-fixed}(1) directly follows
  from Lemma~\ref{lem:size-induct} by choosing $i = 0$, $k = n$.  This
   concludes the proof of Theorem~\ref{lem:size-reg-fixed}(1).

   \section{Complementing regular expressions}
   \label{sec:complement}
   It is known that extended regular expressions are non-elementary more
   succinct than classical ones~\cite{Dang,DBLP:conf/stoc/StockmeyerM73}.
   Intuitively, each exponent in the tower requires nesting of an
   additional complement.  In this section, we show that in defining the
   complement of a single regular expression, a double-exponential size
   increase cannot be avoided in general. In contrast, when the
   expression is one-unambiguous its complement can be computed in
   polynomial time.

   \begin{mytheorem} \label{theo:size-complement}
     \begin{enumerate}
     \item For every regular expression $r$ over $\Sigma$, a
       regular expression $s$ with $L(s)= \Sigma^* \setminus L(r)$ can be
       constructed in time ${\cO}(2^{|r|+1} \cdot |\Sigma| \cdot 4^{2^{|r|+1}})$.

     \item Let $\Sigma$ be a four-letter alphabet. For every
       $n \in \nat$, there is a regular expressions $r_n$ of size
       ${\cO}(n)$ such that any regular expression $r$ defining $\Sigma^*
       \setminus L(r_n)$ is of size at least $2^{2^n}$.
     \end{enumerate}
   \end{mytheorem}
   \begin{myproof}
       (2) Take $\Sigma$ as $\Sigma_K$, that is, $\{0,1,\$,\#\}$.  Let $n
       \in \nat$. We define an expression $r_n$ of size ${\cO}(n)$, such
       that $\Sigma^* \setminus L(r_n) = K_{2^n}$. By
       Theorem~\ref{lem:size-reg-fixed}, any regular expression defining
       $K_{2^n}$ is of size exponential in $2^n$, that is, of size
       $2^{2^n}$.  By $r^{[0,n-1]}$ we abbreviate the expression
       $(\varepsilon + r(\varepsilon + r(\varepsilon \cdots (\varepsilon
       + r))))$, with a \emph{nesting depth} of $n-1$. 
       We then define $r_n$ as the disjunction of the following
       expressions:
       \begin{itemize}
       \item all strings that do not start with a prefix in $(0+1)^n\$$:
	   $$\Sigma^{[0,n]} + (0+1)^{[0,n-1]}(\$ + \#)\Sigma^* + (0+1)^n(0 + 1 + \#)\Sigma^*$$

	 \item all strings where a $\$$ is not followed by a string in
	   $(0+1)^n\#$:
   $$\Sigma^*\$\big(\Sigma^{[0,n-1]}(\# + \$) +
   \Sigma^n(0+1+\$)\big)\Sigma^*$$

   \item all strings where a non-final $\#$ is not followed by a string
     in $(0+1)^n\$$:
   $$\Sigma^*\#\big(\Sigma^{[0,n-1]}(\# + \$) +
   \Sigma^n(0+1+\#)\big)\Sigma^*$$

   \item all strings that do not end in $\#$:
   $$\Sigma^*(0+1+\$)$$

   \item all strings where the corresponding bits of corresponding blocks
     are different:
   $$((0+1)^* + \Sigma^*\#(0+1)^*)0\Sigma^{3n+2}1\Sigma^* +
   ((0+1)^* + \Sigma^*\#(0+1)^*)1\Sigma^{3n+2}0\Sigma^*.$$
   \end{itemize}
   It should be clear that a string over $\{0,1,\$,\#\}$ is matched by
   none of the above expressions if and only if it belongs to $K_{2^n}$.
   So, the complement of $r_n$ defines exactly $K_{2^n}$.
   \end{myproof}

   The previous theorem essentially shows that in complementing a regular
   expression, there is no better algorithm than translating to a DFA,
   computing the complement and translating back to a regular expression
   which includes two exponential steps.  However, when the given regular
   expression is one-unambiguous, a corresponding DFA can be computed in
   quadratic time through the Glushkov
   construction~\cite{Bruggemann-KleinW1998-dup} eliminating already one
   exponential step. The proof of the next theorem shows that the
   complement of that DFA can be directly defined by a regular expression
   of polynomial size.

   \begin{mytheorem} \label{theo:comp-uab-poly} For any one-unambiguous
     regular expression $r$ over an alphabet $\Sigma$, a regular
     expression $s$ defining $\Sigma^* \setminus L(r)$ can be constructed
     in time $\cO(n^3)$, where $n$ is the size of $r$.
   \end{mytheorem}

   \begin{myproof}
     Let $r$ be a one-unambiguous expression over $\Sigma$.  We introduce
     some notation.
     \begin{itemize}
     \item The set $\nfirst(r)$ contains all $\Sigma$-symbols which are
       not the first symbol in any word defined by $r$, that is,
       $\nfirst(r) = \Sigma \setminus \{a \mid a \in \Sigma \wedge
       \exists w \in \Sigma^*, aw \in L(r)\}$ .
     \item For any symbol $x \in \sym(r^\flat)$, the set $\nfollow(r,x)$
       contains all $\Sigma$-symbols of which no marked version can
       follow $x$ in any word defined by $r^\flat$. That is,
       $\nfollow(r,x) = \Sigma \setminus \{y^\natural \mid y \in
       \sym(r^\flat) \wedge \exists w,w' \in \sym(r^\flat)^*, wxyw' \in
       L(r^\flat)\}$.
     \item The set $\last(r)$ contains all \emph{marked} symbols which
       are the last symbol of some word defined by $r^\flat$. Formally,
       $\last(r) = \{x \mid x \in \sym(r^\flat) \wedge \exists w \in
       \Sigma^*, wx \in L(r^\flat)\}$.
     \end{itemize}
     We define the following regular expressions:
     \begin{itemize}
     \item $\init(r) = \left\{\begin{array}{ll}
	   \nfirst(r)\Sigma^* & \text{if $\varepsilon \in L(r)$; and}\\
	   \varepsilon + \nfirst(r)\Sigma^* &\text{if $\varepsilon \notin
	     L(r)$}.
	 \end{array}\right.$

     \item For every $x \in \sym(r^\flat)$, let $r^\flat_x$ be the
       expression defining $\{wx \mid w \in \sym(r^\flat)^* \wedge
       \exists u \in \sym(r^\flat)^*, wxu \in L(r^\flat)\}$. That is, all
       prefixes of strings in $r^\flat$ ending in $x$. Then, let $r_x$
       define $L(r^\flat_x)^\natural$.
     \end{itemize}
     We are now ready to define $s$:
     \begin{multline*}
       \init(r) + \bigcup_{x \notin \last(r)} r_x(\varepsilon +
       \nfollow(r,x)\Sigma^*) + \bigcup_{x \in \last(r)}
       r_x\nfollow(r,x)\Sigma^*.
     \end{multline*}
     It can be shown that $s$ can be constructed in time cubic in the
     size of $r$ and that $s$ defines the complement of $r$.  The latter
     is proved by exhibiting a direct correspondence between $s$ and the
     complement of the Glushkov automaton of $r$. \end{myproof}




   We conclude this section by remarking that one-unambiguous regular
   expressions are not closed under complement and that the constructed
   $s$ is therefore not necessarily one-unambiguous.

   \section{Intersecting regular expressions}

   In this section, we study the succinctness of intersection.  In
   particular, we show that the intersection of two (or any fixed number)
   and an arbitrary number of regular expressions are exponentially and
   double exponentially more succinct than regular expressions,
   respectively.  Actually, the exponential bound for a fixed number of
   expressions already holds for single-occurrence regular expressions,
   whereas the double exponential bound for an arbitrary number of
   expressions only carries over to one-unambiguous expressions. For
   single-occurrence expressions this can again be done in exponential
   time.

   In this respect, we introduce a slightly altered version of $K_n$.
   \begin{mydefinition} \label{def:K-F} Let $\Sigma_L =
     \{0,1,\$,\#,\clubsuit\}$. For all $n \in \nat$, $L_n = \{\rho_n(w)\clubsuit \mid w \in Z_n \wedge
     |w| \text{ is even}\}$.
   \end{mydefinition}
   We also define a variant of $Z_n$ which only slightly alters the
   $a_{i,j}$ symbols in $Z_n$. Thereto, let
   $\Sigma_n^\circ=\{a_{i^{\circ},j},a_{i,j^\circ} \mid 0 \leq i,j < n\}$
   and set $\hat\rho(a_{i,j}a_{j,k}) = \triangleOne_i a_{i,j^\circ}a_{j^\circ,k}$ and 
    $\hat\rho(
   a_{i_0,i_1}a_{i_1,i_2}\allowbreak \cdots a_{i_{k-2},i_{k-1}}a_{i_{k-1},i_k})=
   \hat\rho(a_{i_0,i_1}a_{i_1,i_2})\cdots
   \hat\rho(a_{i_{k-2},i_{k-1}}a_{i_{k-1},i_k})$, where $k$ is even.
   \begin{mydefinition} \label{def:K-G} Let $n \in \nat$ and $\Sigma_M^n
     =  \Sigma_n^\circ \cup \{\triangleOne_0,\triangleTwo_0,\ldots,\triangleOne_{n-1},\triangleTwo_{n-2}\}$. Then, $M_n =
     \{\hat\rho(w)\triangleTwo_i \mid w \in Z_n \wedge |w| \text{ is
       even} \wedge i \text{ is the end-point of } w\}$.
   \end{mydefinition}
   Note that paths in $M_n$ are those in $Z_n$ where every odd position
   is promoted to a circled one $(^\circ)$, and triangles labeled with
   the non-circled positions are added. For instance, the string
   $a_{2,4}a_{4,3}a_{3,3}a_{3,0} \in Z_5$ is mapped to the string
   $\triangleOne_2a_{2,4^\circ}a_{4^\circ,3}\triangleOne_3a_{3,3^\circ}a_{3^\circ,0}\triangleTwo_0
   \in M_5$.



   We make use of the following property: 
   \begin{mylemma}  \label{lem:size-reg-fixed-FG}
     Let $n \in \nat$.
     \begin{enumerate}
     \item 
   \label{lem:size-reg-fixed-F} Any
     regular expression defining $L_n$ is of size at least
     $2^n$.
   \item 
     \label{lem:size-reg-fixed-G} Any regular expression defining $M_n$
     is of size at least $2^{n-1}$.
   \end{enumerate}
   \end{mylemma}

   The next theorem shows the succinctness of the intersection
   operator.
   \begin{mytheorem} \label{theo:size-intersection-two}
     \begin{enumerate}
     \item For any $k \in \nat$ and regular expressions $r_1,\ldots,r_k$,
       a regular expression defining $\bigcap_{i \leq k} L(r_k)$ can be
       constructed in time ${\cO}((m+1)^k \cdot |\Sigma| \cdot
       4^{(m+1)^k})$, where $m = \max{\{|r_i| \mid 1 \leq i \leq k\}}$.

     \item For every $n \in \nat$, there are SOREs $r_n$ and $s_n$ of
       size $\cO(n^2)$ such that any regular expression defining $L(r_n)
       \cap L(s_n)$ is of size at least $2^{n-1}$.

     \item For each $r \in \rein$ an equivalent regular expression can be
       constructed in time ${\cO}(2^{|r|} \cdot |\Sigma| \cdot 4^{2^{|r|}})$.

     \item For every $n \in \nat$, there are one-unambiguous regular
       expressions $r_1, \ldots, r_m$, with $m=2n+1$, of size ${\cO}(n)$
       such that any regular expression defining $\bigcap_{i \leq m}
       L(r_i)$ is of size at least $2^{2^n}$.

       \item  Let $r_1,\ldots,r_n$ be SOREs. A
       regular expression defining $\bigcap_{i \leq n} L(r_n)$ can be constructed in
       time ${\cO}(m \cdot |\Sigma| \cdot 4^{m})$, where $m =
       \sum_{i \leq n} |r_i|$.

    \end{enumerate}
   \end{mytheorem}
   \begin{myproof}

   \noindent
   (2) Let $n \in \nat$. By
   Lemma~\ref{lem:size-reg-fixed-FG}(\ref{lem:size-reg-fixed-G}), any
   regular expression defining $M_{n}$ is of size at least
   $2^{n-1}$.  We define SOREs $r_n$ and $s_n$ of size quadratic in $n$,
   such that $L(r_n) \cap L(s_n) = M_n$. We start by partitioning
   $\Sigma_M^n$ in two different ways.  To this end, for every $i < n$, define
   $\Out_i=\{a_{i,j^\circ} \mid 0 \leq j < n\}$, $\In_i=\{a_{j^\circ,i}
   \mid 0 \leq j < n\}$, $\Out_{i^\circ}=\{a_{i^\circ,j} \mid 0 \leq j <
   n\}$, and, $\In_{i^\circ}=\{a_{j,i^\circ} \mid 0 \leq j < n\}$.  Then,
   $$\Sigma_M^n=\bigcup_{i} \In_i\cup\Out_i\cup\{\triangleOne_i,
   \triangleTwo_i\} = \bigcup_{i^\circ}
   \In_{i^\circ}\cup\Out_{i^\circ}\cup\{\triangleOne_i,
   \triangleTwo_i\}.$$
   Further, define $$r_n = ((\triangleOne_0 + \cdots + \triangleOne_{n-1}) \bigcup_{i^\circ} \In_{i^\circ}
   \Out_{i^\circ})^+(\triangleTwo_0 + \cdots + \triangleTwo_{n-1})$$ and $$s_n = \Bigl(\bigcup_i
   (\In_{i}+\varepsilon)(\triangleOne_i + \triangleTwo_i)(\Out_{i}+
   \varepsilon)\Bigr)^*.$$ 

   Now, $r_n$ checks that every string consists
   of a sequence of blocks of the form $\triangleOne_i
   a_{j,k^\circ}a_{k^\circ,\ell}$, for $i,j,k,\ell < n$, ending with a
   $\triangleTwo_i$, for $i < n$. It thus sets the format of the strings and checks
   whether the circled indices are equal. Further, $s_n$ checks whether
   the non-circled indices are equal and whether the triangles have the
   correct indices. Since the alphabet of $M_n$ is of
   size $\cO(n^2)$, also $r_n$ and $s_n$ are of size $\cO(n^2)$.



   \noindent
   (4) Let $n \in \nat$. We define $m=2n+1$ one-unambiguous regular
   expressions of size ${\cO}(n)$, such that their intersection defines
   $L_{2^n}$. By Lemma~\ref{lem:size-reg-fixed-FG}(\ref{lem:size-reg-fixed-F}), any regular expression
   defining $L_{2^n}$ is of size at least $2^{2^n}$ and the theorem
   follows.  For ease of readability, we denote $\Sigma_L$ simply by
   $\Sigma$. The expressions are as follows. There should be an even length sequence of blocks:
    $$\big((0+1)^n\$(0+1)^n\#(0+1)^n\$(0+1)^n\#\big)^*\spadesuit.$$ 
   For all $i\in \{0,\ldots,n-1\}$,
    the $(i+1)$th bit of the two numbers surrounding an odd $\#$
       should be equal:
   $$\big(\Sigma^i(0\Sigma^{3n + 2}0 + 1\Sigma^{3n +
     2}1)\Sigma^{n-i-1}\#\big)^*\spadesuit.$$
   For all $i\in \{0,\ldots,n-1\}$,
   the $(i+1)$th bit of the two numbers surrounding an even $\#$ should
     be equal:
   \begin{eqnarray*}
   \Sigma^{2n+2}\Big(\Sigma^i
     (0\Sigma^{2n-i+1}(\spadesuit +
     \Sigma^{n+i+1}0\Sigma^{n-i-1}\#)
    + (1\Sigma^{2n-i+1}(\spadesuit +
     \Sigma^{n+i+1}1\Sigma^{n-i-1}\#))) \Big)^*.
   \end{eqnarray*}
   Clearly, the intersection of the above expressions defines $L_{2^n}$.
   Furthermore, every expression is of size $\cO(n)$ and is
   one-unambiguous as the Glushkov construction translates them into
   a DFA~\cite{Bruggemann-KleinW1998-dup}. \end{myproof}

\vskip-.3cm
   \section{Conclusion}
   In this paper we showed that the complement and intersection of regular
   expressions are double exponentially more succinct than ordinary
   regular expressions. For complement, complexity can be reduced to
   polynomial for the class of one-unambiguous regular expressions
   although the obtained expressions could fall outside that class.  For
   intersection, restriction to SOREs reduces complexity to exponential.
   It remains open whether there are natural classes of regular
   expressions for which both the complement and intersection can be
   computed in polynomial time.

\bigskip
\noindent {\it Acknowledgment.} We thank Juraj Hromkovi$\check{\text{c}}$ for sending
us reference~\cite{waizenegger}.  

\end{document}